\documentclass[sigconf]{acmart}

\usepackage{multirow}
\usepackage{color, colortbl}
\usepackage{arydshln}
\usepackage{array}
\usepackage{balance} 
\usepackage{subcaption}
\usepackage{balance} 

\definecolor{Gray}{gray}{0.9}
\definecolor{ashgrey}{rgb}{0.7, 0.75, 0.71}

\newcommand{\xmind}{xMIND}

\newcommand{\zsxlt}{\texttt{ZS-XLT}}
\newcommand{\fsxlt}{\texttt{FS-XLT}}

\newcommand{\zsxltmono}{\texttt{ZS-XLT\textsubscript{MONO}}}
\newcommand{\zsxltbiling}{\texttt{ZS-XLT\textsubscript{BILING}}}
\newcommand{\fsxltmono}{\texttt{FS-XLT\textsubscript{MONO}}}
\newcommand{\fsxltbiling}{\texttt{FS-XLT\textsubscript{BILING}}}

\AtBeginDocument{%
  \providecommand\BibTeX{{%
    \normalfont B\kern-0.5em{\scshape i\kern-0.25em b}\kern-0.8em\TeX}}}

\copyrightyear{2024}
\acmYear{2024}
\setcopyright{rightsretained}
\acmConference[SIGIR '24] {Proceedings of the 47th International ACM SIGIR Conference on Research and Development in Information Retrieval}{July 14--18, 2024}{Washington, DC, USA.}
\acmISBN{979-8-4007-0431-4/24/07}
\acmDOI{XXXXXXX.XXXXXXX}

\begin{document}

\title{MIND Your Language: A Multilingual Dataset for Cross-lingual News Recommendation}

\author{Andreea Iana}
\affiliation{%
  \institution{University of Mannheim}
  \city{Mannheim}
  \country{Germany}
}
\email{andreea.iana@uni-mannheim.de}

\author{Goran Glavaš}
\affiliation{%
  \institution{CAIDAS, University of Würzburg}
  \city{Würzburg}
  \country{Germany}}
\email{goran.glavas@uni-wuerzburg.de}

\author{Heiko Paulheim}
\affiliation{%
  \institution{University of Mannheim}
  \city{Mannheim}
  \country{Germany}
}
\email{heiko.paulheim@uni-mannheim.de}

\renewcommand{\shortauthors}{Andreea Iana, Goran Glavaš, \& Heiko Paulheim}

\begin{abstract}
Digital news platforms use news recommenders as the main instrument to cater to the individual information needs of readers. 
Despite an increasingly language-diverse online community, in which many Internet users consume news in multiple languages, the majority of news recommendation focuses on major, resource-rich languages, and English in particular. Moreover, nearly all news recommendation efforts assume \textit{monolingual} news consumption, whereas more and more users tend to consume information in at least two languages. Accordingly, the existing body of work on news recommendation suffers from a lack of publicly available multilingual benchmarks that would catalyze development of news recommenders effective in multilingual settings and for low-resource languages. 
Aiming to fill this gap, we introduce \xmind{}, an \textit{open, multilingual} news recommendation dataset derived from the English MIND dataset using machine translation, covering a set of 14 linguistically and geographically diverse languages, with digital footprints of varying sizes. 
Using \xmind{}, we systematically benchmark several state-of-the-art content-based neural news recommenders (NNRs) in both zero-shot (\zsxlt{}) and few-shot (\fsxlt{}) cross-lingual transfer scenarios, considering both monolingual and bilingual news consumption patterns. Our findings reveal that (i) current NNRs, even when based on a multilingual language model, suffer from substantial performance losses under \zsxlt{} and that (ii) inclusion of target-language data in \fsxlt{} training has limited benefits, particularly when combined with a bilingual news consumption. Our findings thus warrant a broader research effort in multilingual and cross-lingual news recommendation. The \xmind{} dataset is available at \href{{https://github.com/andreeaiana/xMIND}}{https://github.com/andreeaiana/xMIND}.

\end{abstract}



\maketitle

\section{Introduction}
\label{sec:intro}

The digitalization of news consumption has established news platforms as the prevalent medium of information for Internet users. This, in turn, propelled personalized news recommendation systems into the main vehicle used by news websites to cater the individual information needs of readers. 
The global expansion of the Internet's outreach has vastly increased the language diversity of its users \cite{zuckerman2008,ling2020multilingual}, a non-negligible fraction of whom are polyglots, speaking and consuming news in two or more languages. For example, 22\% of Americans speak a language other than English at home\footnote{\url{https://data.census.gov/table/ACSST1Y2022.S1601?q=language}}, whereas 65\% of the working-age adults in the European Union know at least one foreign language.\footnote{\url{https://ec.europa.eu/eurostat/web/products-eurostat-news/-/EDN-20180926-1}} 
However, the majority of online content remains available mostly in few resource-rich languages, with English accounting for more than half of the digital texts, while the majority of languages spoken around the globe appear in less than 0.1\% of the websites or are not represented online.\footnote{\url{https://w3techs.com/technologies/overview/content_language}}

News media play a central role in democratic societies, by ensuring informed citizens and providing a public forum for disseminating and debating ideas and opinions \cite{balkin2017free,helberger2021democratic}. In this context, recommender systems shape people's worldviews and opinions through the way in which they filter and propagate news \cite{pariser2011filter}. On the one hand, research on personalized neural news recommenders \cite{wu2023personalized} has focused on improving their accuracy and diversity \cite{wu2020sentirec,wu2022end,wu2022removing,shi2022dcan}, by mitigating technical challenges in news encoding \cite{wang2018dkn,wu2019npa,wu2019nrms,wu2019naml,wu2019tanr,liu2020kred,qi2021personalized,wu2021empowering,yu2022tiny,jiang2023tadi,li2023pbnr,zhang2023prompt} and user modeling \cite{an2019lstur,qi2021pp,qi2021hierec,qi2022news,wang2022news,li2022miner,qi2022fum,iana2023simplifying}. On the other hand, recent progress in areas such as neural machine translation (NMT) \cite{dabre2020survey,fan2021beyond,haddow2022survey,costa2022no,kudugunta2023madlad} and multilingual large language models (mPLMs) \cite{conneau2019cross,conneau2020unsupervised,wei2020learning,xue2021mt5,workshop2022bloom,touvron2023llama,wei2023polylm} for low(er)-resource languages has begun to democratize access to information for underrepresented communities \cite{joshi2020state}.

Nonetheless, the existing body of research on news recommendation is limited in two main dimensions. Firstly, there is a \textit{scarcity of publicly-available, diverse, multilingual news recommendation datasets} that could be leveraged to develop efficient multilingual news recommendation and support effective cross-lingual transfer to resource-lean languages. Despite the availability of adequate datasets being paramount to developing high-quality recommenders (e.g., see 
the Amazon dataset\footnote{\url{https://cseweb.ucsd.edu/~jmcauley/datasets.html\#amazon_reviews}} for product recommendation, or MovieLens \cite{harper2015movielens} and Netflix \cite{bennett2007netflix} for movie recommendation),
the vast majority of news recommendation benchmarks are monolingual \cite{kille2013plista,gulla2017adressa, de2018news,gabriel2019contextual,lucas2023npr,wu2020mind}. Furthermore, the few existent multilingual benchmarks comprise only high-resource languages, and are either topic-specific and small-sized \cite{iana2023nemig} or proprietary \cite{wu2021empowering}. Secondly, the design of news recommenders for multi- and cross-lingual settings has been left largely unexplored. Traditional news recommendation systems support mostly monolingual recommendations, thus hindering simultaneous browsing for news across multiple languages. In practice, this translates into less relevant, less balanced and less diverse recommendations for multilingual news consumers \cite{ling2020multilingual}. Overall, these limitations pose significant challenges for online news readers who are multilingual or consume content in resource-lean and/or underrepresented languages. 

\vspace{1.4mm} \noindent \textbf{Contributions.}
We address the above research gaps by introducing \xmind{}, a new public and large-scale multilingual news dataset for multi- and cross-lingual recommendation. \xmind{} is derived by translating the articles of the English-only dataset MIND \cite{wu2020mind} into a diverse selection of 14 high- and low-resource languages, spanning five geographical macro-areas and 13 distinct language families using the open-source machine translation system NLLB \cite{costa2022no}. Compared to existing multilingual news recommendation datasets, \xmind{} is: \textbf{(1)} much more \textit{diverse} -- we include both resource-rich and resource-poor languages, covering a wide variety of geographical areas, family languages, and scripts, with some of the languages being out-of-sample for existing mPLMs (i.e., not present in the mPLM's pretraining); \textbf{(2)} \textit{parallel} -- the same set of news has been translated into all target languages, enabling the direct comparison of the performance of multilingual news recommenders and cross-lingual transfer approaches across target languages; \textbf{(3)} \textit{open source} -- we release the dataset in the TSV format 
and provide scripts for loading and combining the news with the corresponding click logs from the MIND dataset in the NewsRecLib \cite{iana2023newsreclib} library. 

We use \xmind{} to benchmark a range of state-of-the-art neural news recommenders, in zero-shot (\zsxlt{}) and few-shot (\fsxlt{}) cross-lingual transfer setups, covering two realistic news consumption scenarios: monolingual and bilingual. We show that recommenders trained monolingually on English news suffer significant performance drops when evaluated on the target languages under \zsxlt{}, even when paired with a massively multilingual language model. More importantly, we demonstrate that target-language injection during training has a limited effect in mitigating these performance drops, revealing the urgent need for developing more accurate and robust cross-lingual news recommendation approaches. Lastly, we investigate the quality of the translations in \xmind{} through a human-based annotation task and comparison against translations obtained with a commercial NMT system.

\section{Related Work}

\vspace{1.4mm} \noindent \textbf{News Recommendation.}
Personalized news recommendation aims to alleviate the information overload of online news readers by providing suggestions tailored to their individual preferences \cite{li2019survey,wu2023personalized}. Neural news recommenders (NNRs) have become the driver of personalized news recommendation, replacing systems relying on manual feature engineering \cite{wu2023personalized}. The majority of NNRs commonly consist of a dedicated (i) news encoder (NE), (ii) user encoder (UE), and a (iii) click predictor. The NE learns news representations from various input features (e.g., title, topical categories, named entities), either by instantiating convolutional neural networks \cite{wang2018dkn,wu2019naml,wu2019tanr}, self-attention networks \cite{wu2019nrms,qi2021pp}, or graph attention networks \cite{qi2021personalized} with pretrained word embeddings, or, more recently, by leveraging pretrained language models \cite{wu2021empowering,yu2022tiny,zhang2023prompt,li2023pbnr}. Afterwards, the UE aggregates and contextualizes the embeddings of a user's clicked news into a user-level representation by means of sequential \cite{an2019lstur,qi2020privacy,wang2022news} or attentive \cite{wu2019naml,wu2019nrms} encoders. Lastly, a candidate article's recommendation score is computed by comparing its embedding against the user profile \cite{wu2023personalized}. Although a significant body of work has sought to improve personalization by enhancing NNRs' core components -- news and user modeling -- the vast majority of efforts have been nearly exclusively deployed in monolingual settings. More specifically, despite the abundance of polyglot news readers, few works explore the behavior of NNRs in a multi- or cross-lingual scenario. \cite{wu2021empowering} suggested instantiating the NE with mPLMs to enable news recommendation in diverse languages. \citet{guo2023few} proposed a new NE based on an unsupervised cross-lingual transfer model to address the few-shot recommendation problem between record-rich and unpopular or early-stage recommendation platforms without overlapping users and with news in different languages. However, these works  focus exclusively on (i) resource-rich languages and (ii) monolingual news consumption. News recommendation for multilingual news consumers, especially speakers of under-resourced languages, thus remains largely unexplored.

\vspace{1.4mm} \noindent \textbf{Recommendation Datasets.}
The advancement of recommender systems heavily depends on the existence and availability of suitable datasets. In the past decade, several public monolingual datasets have been proposed for training and benchmarking news recommenders: Plista \cite{kille2013plista} (German), Adressa \cite{gulla2017adressa} (Norwegian), Globo \cite{de2018news,gabriel2019contextual} and its recent improved version NPR \cite{lucas2023npr} (Portuguese), and MIND \cite{wu2020mind} (English). Among these, the MIND dataset has become a reference benchmark for the news recommendation community, given the limitations of the earlier datasets, such as a lack of original news texts, metadata information, or limited dataset size \cite{wu2020mind}. However, these datasets consists only of monolingual news, and therefore, hinder the development of multilingual recommender systems. \citet{iana2023nemig} aimed to address this problem by proposing NeMig, a multilingual news recommendation dataset in English and German. NeMig contains articles on the topic or refugees and migration collected from German and US media outlets, and rich user data encompassing both click logs and demographic and political information. Besides covering only two major languages, NeMig is small (7K German and  10K English articles) and covers only one specific topic. \citet{wu2021empowering} mention the multilingual news recommendation dataset collected from the MSN News platform to analyze the effectiveness of mPLM-based NEs in multilingual news recommendations. Their dataset contains user data from seven countries (US, Germany, France, Italy, Japan, Spain, and Korea). Besides all seven included languages being very highly resourced, the dataset is proprietary, i.e., it is not publicly available. 

\section{Dataset Creation}
\label{sec:dataset}

\begin{table*}[t]
    \centering
    \caption{The 14 languages of \xmind{}. We display the language \textit{Code} (ISO 693-3), language name, \textit{Script}, \textit{Macro-area}, and language \textit{Family} and \textit{Genus}. \textit{Res.} indicates whether the language is classified as high or low-resource according to \cite{costa2022no}.}
    \label{tab:languages}
    \resizebox{\textwidth}{!}{%
    \begin{tabular}{llllllrcc}
    \toprule
        \textbf{Code} & \textbf{Language} & \textbf{Script} & \textbf{Macro-area}  & \textbf{Family} & \textbf{Genus} & \textbf{Total Speakers (M)} & \textbf{Res.} & \textbf{mPLM}\\
        \hline
         \texttt{SWH} & Swahili & Latin & Africa & Niger-Congo & Bantu & 71.6 & high & yes \\
         
         \texttt{SOM} & Somali & Latin & Africa & Afro-Asiatic & Lowland East Cushitic & 22.0 & low & yes\\
        
         \texttt{CMN} & Mandarin Chinese & Han & Eurasia & Sino-Tibetan & Sinitic & 1,138.2 & high & yes \\
         
         \texttt{JPN} & Japanese & Japanese & Eurasia & Japonic & Japanesic & 1,234.5 & high & yes \\

        \texttt{TUR} & Turkish & Latin & Eurasia & Altaic & Turkic & 90.0 & high & yes \\
         
         \texttt{TAM} & Tamil & Tamil & Eurasia & Dravidian & Dravidian & 86.6 &  low & yes \\
        
         \texttt{VIE} & Vietnamese & Latin & Eurasia & Austro-Asiatic & Vietic  & 85.8 & high & yes\\
         
         \texttt{THA} & Thai & Thai & Eurasia & Tai-Kadai & Kam-Tai  & 60.8 & high & yes\\
         
         \texttt{RON} & Romanian & Latin & Eurasia & Indo-European & Romance  & 24.5  & high & yes\\
         
         \texttt{FIN} & Finnish & Latin & Eurasia & Uralic & Finnic & 5.6 & high & yes \\
         
         \texttt{KAT} & Georgian & Georgian & Eurasia & Kartvelian & Georgian-Zan  & 3.9 & low & yes\\
        
         \texttt{HAT} & Haitian Creole & Latin & North-America & Indo-European & Creoles and Pidgins & 13.0  & low & no \\
         
         \texttt{IND} & Indonesian & Latin & Papunesia & Austronesian & Malayo-Sumbawan  & 199.1 & high & yes\\
         
         \texttt{GRN} & Guarani & Latin & South America & Tupian & Maweti-Guarani  & (L1 only) 6.7 & low & no\\
         
    \bottomrule
    \end{tabular}%
    }
\end{table*}

We create \xmind{} with two primary considerations in mind: (1) covering languages that are mutually \textit{diverse} linguistically, geographically, and in terms of amount of available text corpora (i.e., high or low resource) and (2) creating a multilingual news dataset that is multi-parallel, i.e., where an article (i.e., a translation thereof) exists in each covered language. The former allows for a more realistic estimate of global multilingual and cross-lingual performance of news recommendation models \cite{joshi2020state,ponti2020xcopa}, whereas the latter enables direct comparability of recommenders' performance across target languages. 
We thus create \xmind{} by translating 130,379 unique news articles from the train, development, and test portions of the English MIND dataset \cite{wu2020mind} (i.e., union of MINDlarge and MINDsmall) into 14 different languages using the NLLB 3.3B open-source NMT model \cite{costa2022no}. The MIND news articles consist of a title and an abstract, and are additionally annotated with the topical category and Wikipedia-disambiguated named entities extracted from the title and abstract.\footnote{Note that 5.4\% of the news in the entire MIND dataset do not contain an abstract.}
Note that, although we translate only the title and abstract of each news, these can still be combined with the corresponding linked named entities for usage in knowledge-aware recommendation models \cite{iana2022survey}.

\vspace{1.4mm} \noindent \textbf{Language Selection.}
We select target languages for \xmind{} based on the following criteria: (1) linguistic diversity in terms of typological properties \cite{wals,littell2017uriel}, language family, and geographical provenance, (2) script diversity, (3) amount of available language resources, primarily raw corpora (i.e., inclusion of both high- and low-resource languages), and (4) coverage by NLLB \cite{costa2022no}. Table \ref{tab:languages} lists the selected languages, summarizing the following information, in accordance with the \#BenderRule \cite{bender2019benderrule}: 

\begin{itemize}
    \item \textbf{Code}: The three-letter ISO 693-3 code of the language;
    \item \textbf{Language}: In case of multiple denominations, we use the language name from the World Atlas of Structures (WALS) \cite{wals}. We cross-reference the names with two other major linguistic resources, Glottolog \cite{hammarstrom2021glottolog} and Ethnologue \cite{lewis2009ethnologue};
    \item \textbf{Script}: We provide the English name of the script; 
    \item \textbf{Family and Genus}: Language family and genus from WALS \cite{wals} and Glottolog \cite{hammarstrom2021glottolog};
    \item \textbf{Resource Level}: We borrow NLLB's \cite{costa2022no} classification of languages into \textit{low-} and \textit{high-resource}; 
    \item \textbf{mPLM Support}: We indicate whether the language is included in the pretraining corpora of XLM-RoBERTa \cite{conneau2020unsupervised}, the representative mPLM used in our experiments (\S\ref{sec:experiments});
    \item \textbf{Total Speakers}: We report the total number of speakers of the language, including L1-level (first-language) and L2-level (second-language) speakers, according to Ethnologue.\footnote{We use the latest statistics available in January 2024 at \url{https://www.ethnologue.com/}.} 
\end{itemize}

\begin{table}
    \centering
    \caption{Indices of typological, genealogical, and geographical diversity for the language samples of different multilingual news recommendation datasets.}
    \label{tab:diversity_indices}
    \resizebox{\columnwidth}{!}{%
    \begin{tabular}{llccc}
    \toprule
         & \textbf{Range} & \textbf{\xmind{}} & \textbf{NeMig}  & \textbf{Wu et al.} \\
        \cmidrule{3-5}

        Typology & [0, 1] & 0.42 & 0.05 & 0.31 \\
        Family & [0, 1] & 0.93 & 0.50 & 0.43 \\
        Geography & [0, ln 6] & 1.13 & 0.00 & 0.00 \\
         
    \bottomrule
    \end{tabular}%
    }
\end{table}

We follow \citet{ponti2020xcopa} and compute three different diversity scores for our language sample: (i) typology index, (ii) family index, and (iii) geographical index. \textbf{1)} The \textit{typology index} is based on 103 typological binary features of each language from URIEL \cite{littell2017uriel}: each feature indicates the presence or absence of a particular linguistic property in a language. 
As per \cite{ponti2020xcopa}, we compute the typology index as the average of entropy scores computed independently for each feature;\footnote{The entropy of a feature for which all languages in the sample have the same value is 0; the entropy has the maximal value ($\log 2$) if the feature is present for the same number of languages as for which it is absent.} 
\textbf{2)} The \textit{family index} is the number of distinct language families divided by the sample size; \textbf{3)} The \textit{geography index} is the entropy of the distribution of languages in the sample over 6 geographic macro-areas of the world.\footnote{The six macro-areas, as defined by \citet{wals}, are: Africa, Australia, Eurasia, North America, Papunesia, and South America.}

Table \ref{tab:diversity_indices} reports the three metrics for \xmind{}, as well as for NeMig \cite{iana2023nemig} and the proprietary dataset from \cite{wu2021empowering} (dubbed \emph{Wu et al.}). \xmind{} offers the most diverse sample in terms of all diversity indices. The sample of languages spans five out of the six macro-areas, and 13 distinct language families covering 14 different genera. We excluded languages from Australia, as they (i) have an extremely low number of native speakers (i.e., at most spoken by a few thousand people), and (ii) are not supported by NLLB \cite{costa2022no}. Additionally, \texttt{GRN} and \texttt{HAT} (i) are spoken in South and North America, both originating from underrepresented macro-areas, and (ii) have not been seen in pretraining of XLM-RoBERTa \cite{conneau2020unsupervised}.
Moreover, \xmind{} covers five low-resource languages (Somali, Tamil, Georgian, Haitian Creole, and Guarani) and six different scripts: Latin and Georgian are \textit{alphabet} scripts; Japanese and Chinese Han are \textit{logographic} scripts, whereas the Tamil and Thai are written in \textit{Abugida} script type.

\vspace{1.4mm} \noindent \textbf{NLLB: Hyperparameter Tuning}
We tune the hyperparameters of the NLLB translation model \cite{costa2022no} using a subset of Global Voices (GV) \cite{TIEDEMANN12.463} as validation set.\footnote{We use the most recent version available online in October 2023, namely \textit{GlobalVoices v2018q4}. The original data can be accessed at \href{https://opus.nlpl.eu/GlobalVoices/corpus/version/GlobalVoices}{https://opus.nlpl.eu/GlobalVoices/corpus/version/GlobalVoices}.}
GV constitutes a parallel corpus of news stories in 46 languages collected from the Global Voices website.\footnote{\href{https://globalvoices.org/}{https://globalvoices.org/}}
We construct the validation dataset by selecting the data files for all covered pairs of \textit{English} (\texttt{ENG}) as the source and any of the \xmind{} languages of \xmind{} as the target: this results in six language pairs, statistics of which are reported in Table \ref{tab:gobal_voices_stats}.

\begin{table}[t]
    \centering
    \caption{Statistics of the subset of Global Voices \cite{TIEDEMANN12.463} used as validation data for tuning the NLLB \cite{costa2022no} hyperparameters.}
    \label{tab:gobal_voices_stats}
    \begin{tabular}{lrr}
    \toprule
        \textbf{Language Pair} & \textbf{Sentence pairs} & \textbf{Words (M)}  \\
        \hline
        
        \texttt{ENG} -> \texttt{CMN} & 137,737 & 2.83 \\
        \texttt{ENG} -> \texttt{SWH} & 30,338 & 1.13 \\
        \texttt{ENG} -> \texttt{IND} & 15,266 & 0.54 \\
        \texttt{ENG} -> \texttt{JPN} & 8,595 & 0.18 \\
        \texttt{ENG} -> \texttt{TUR} & 7,479 & 0.24 \\
        \texttt{ENG} -> \texttt{RON} & 4,265 & 0.17\\
         
    \bottomrule
    \end{tabular}%
\end{table}

\begin{table}[t]
    \centering
     \caption{Hyperparameter optimization results on the subset of Global Voices \cite{TIEDEMANN12.463}. We report only the results obtained with the best number of beams. We report macro-average sacreBLEU scores over six language pairs.}
    \label{tab:nllb_hpo_results}
    \begin{tabular}{lcc}
    \toprule
         \textbf{Decoding Strategy} & \textbf{\# Beams} & \textbf{sacreBLEU}\\ \hline

        Greedy & 1 & 18.42 \\
        Multinomial sampling & 1 & 11.97 \\
        Beam Search & 4 & 19.03 \\
        Beam Search Multinomial Sampling & 4  & 18.87 \\
    \bottomrule
    \end{tabular}
\end{table}

We compare four decoding strategies: greedy, multinomial sampling, beam-search, and beam search with multinomial sampling. For the beam-search decoding strategies, we search for optimal number of beams in the range $[2, 8]$ and use default values for all other hyperparameters. We evaluate the translation quality using the sacreBLEU score \cite{post-2018-call}. With the goal of finding the best decoding strategy for a broad range of languages, we compute the macro-average over the six language pairs in our validation dataset. As shown in Table \ref{tab:nllb_hpo_results}, we identify beam search decoding with 4 beams as the best choice: using more beams increases computational cost while bringing negligible sacreBLEU gains. 

\vspace{1.4mm} \noindent \textbf{Final Dataset.}
The \xmind{} dataset contains 130,379 unique news in the 14 different languages listed in Table \ref{tab:languages}. Each article contains a news ID, a translated title, and a translated abstract -- if one was provided in the corresponding English article from MIND \cite{wu2020mind}. Following \citet{wu2020mind}, we split \xmind{}, for each language, firstly into a small and a large version of the dataset, and secondly, into train, development, and test portions, each corresponding to the original splits of the MIND dataset.\footnote{\url{https://msnews.github.io/}}
We release \xmind{} publicly, in tab-separated format at  \url{https://github.com/andreeaiana/xMIND}. \xmind{} can be combined with additional news and behavioral information provided in MIND \cite{wu2020mind}, using the news IDs. Additionally, to facilitate a seamless integration with existing NNRs, we implement the data loading functionality for \xmind{} in NewsRecLib \cite{iana2023newsreclib}. 

\begin{table}[t]
    \centering
    \caption{Number of news in the different splits of \xmind{}.}
    \label{tab:xmind_dataset}
    \begin{tabular}{ccccc}
    \toprule
         \multicolumn{2}{c}{\textbf{Small}} & \multicolumn{3}{c}{\textbf{Large}} \\ \cmidrule(lr){1-2} \cmidrule(lr){3-5}
         Train & Dev & Train & Dev &  Test \\ \hline
         51,282 & 42,416 & 101,527 & 72,023 & 120,959 \\      
         
    \bottomrule
    \end{tabular}%
\end{table}

\section{Experimental Setup}
\label{sec:experiments}

We systematically benchmark a range of state-of-the-art content-based NNRs in zero-shot (\zsxlt{}) and few-shot (\fsxlt{}) cross-lingual transfer scenarios. Our experiments encompass two types of news consumption patterns: monolingual and bilingual.

\subsection{Benchmarked Recommenders}
\vspace{1.4mm} \noindent \textbf{Neural News Recommenders.} 
We evaluate several content-based NNRs: (1) \textit{NAML} \cite{wu2019naml}, (2) \textit{LSTUR} \cite{an2019lstur}, (3) \textit{MINS} \cite{wang2022news}, (4) \textit{CAUM} \cite{qi2022news}, (5) \textit{TANR} \cite{wu2019tanr}, (5) \textit{MINER} \cite{li2022miner}, and (6) \textit{MANNeR} \cite{iana2023train} (only the \texttt{CR-Module} responsible for pure content-based recommendation, without any aspect-based personalization or diversification). Additionally, we use as baseline (7) \textit{NAML\textsubscript{CAT}}, a language-agnostic variant of NAML which learns news embeddings solely based on randomly-initialized category vectors, and user representations by attending over the embeddings of the clicked news.
With the exception of MINER and MANNeR, designed with a PLM-based NE, the remaining NNRs originally contextualize word embeddings with convolutional neural networks (CNNs) \cite{kim-2014-convolutional}, additive attention \cite{bahdanau2014neural}, or multi-head self-attention \cite{vaswani2017attention} networks. For fair comparison and to enable multilingual recommendations, we follow \citet{wu2021empowering}, and replace the original NEs of these NNRs with an mPLM. NAML, LSTUR, MINS, TANR, and CAUM leverage category information in addition to the news text, whereas CAUM and MANNeR also encode named entities. Models with multiple input features either concatenate (i.e, LSTUR, CAUM) or attend over them (i.e, NAML, MINS, MANNeR) to produce the final news embedding.

The recommenders further differ in their UE component: NAML \cite{wu2019naml} and TANR \cite{wu2019tanr} encode users' preferences using additive attention; MINS \cite{wang2022news} combines multi-head self-attention with a multi-channel GRU-based \cite{cho2014learning} recurrent network and additive attention; MINER \cite{li2022miner} introduces a poly-attention approach based on multiple additive attentions to learn various interest vectors for each user. Moreover, LSTUR and CAUM differentiate between short and long-term user preferences. More specifically, LSTUR encodes the former from the clicked news embeddings with a GRU, and the latter via randomly initialized and fine-tuned embeddings; the final user representation is produced by combining the two embeddings  \cite{an2019lstur}.\footnote{Note that in our experiments, we use the \textit{ini} strategy of LSTUR for obtaining the final user embedding, as it outperforms the \textit{con} variant in preliminary evaluations. We refer the reader to \cite{an2019lstur} for more details.}
CAUM models long-term dependencies between clicked news with a candidate-aware self-attention network, short-term user interests from adjacent clicks with a candidate-aware CNN, and obtains the final candidate-aware user embedding by attending over the two intermediate representations \cite{qi2022news}. In contrast to the other models, MANNeR does not learn a parameterized UE, instead using a late fusion approach consisting of the mean-pooling of dot-product scores between each of the candidates and the clicked news \cite{iana2023train}. 

All benchmarked models compute the recommendation score as the dot product between the representations of the candidate and the clicked news. MANNeR is optimized using a supervised contrastive loss (SCL) \cite{khosla2020supervised}, whereas the other models  are trained by minimizing the standard cross-entropy loss. 

\vspace{1.4mm} \noindent \textbf{Data.}
We combine the \xmind{} news with the corresponding click logs and additional news annotations (i.e., categories and named entities) from MIND based on the news IDs. We conduct all experiments on the \textit{small variant} of the resulting dataset. Since \citet{wu2020mind} do not release test labels for MIND, we use the validation portion for testing, and split the respective training set into temporally disjoint training (first four days) and validation (last day) sets. 

\vspace{1.4mm} \noindent \textbf{Training Details.}
We use XLM-RoBERTa \cite{conneau2020unsupervised} as the mPLM in all models, and fine-tune only its last four layers.\footnote{In the interest of computational efficiency, we keep the bottom eight layers of the Transformer encoder frozen.}
We use 100-dimensional TransE embeddings \cite{bordes2013translating}, pretrained on Wikidata, to initialize the entity encoder in the NE of the knowledge-aware models.\footnote{The entity vectors are provided as part of the original MIND dataset \cite{wu2020mind}.}
In line with prior work, we set the maximum history length to 50 and sample four negatives per positive sample during training, as per \citet{ijcai2022infonce}. To ensure comparability, we train all models with mixed precision using the NewsRecLib\footnote{\url{https://github.com/andreeaiana/newsreclib}} library \cite{iana2023newsreclib},  with a batch size of 8, for 10 epochs with early stopping, optimizing with the Adam algorithm \cite{kingma2014adam}. 

We perform hyperparameter optimization for the most important hyperparameters of each NNR using the English news as training and validation sets (i.e., hyperparameter tuning on the MIND dataset). Concretely, we search for the optimal learning rate in the range $[1e^{-3}, 1e^{-4}, 1e^{-5}]$ for all models. We optimize the number of heads in the multi-head self-attention networks of NAML, LSTUR, MINS, TANR, and CAUM in $[8, 12, 16, 24, 32]$, and the query vector dimensionality in the additive attention network in $[50, 200]$ with a step of 50 for NAML, LSTUR, MINS, and TANR. Moreover, we search the optimal SCL temperature in MANNeR sweeping the interval $[0.1, 0.5]$, with a step of 0.02. Lastly, for MINER, we optimize the number of context codes in the interval $[8, 16, 32, 48]$, and choose the best-performing aggregation type between \textit{mean}, \textit{max}, and \textit{weighted}. We set all remaining model-specific hyperparameters to the optimal values reported in the respective papers. We repeat each experiment three times, with different random seeds, and report averages and standard deviations for the standard metrics: AUC, MRR, nDCG@5, and nDCG@10.\footnote{We train all models on one NVIDIA Tesla A100 (with 40/80 GB memory) or A40 (with 48 GB memory).}

\subsection{Cross-Lingual Recommendation Scenarios}
We benchmark the NNRs in two evaluation setups: (i) \textbf{zero-shot (\zsxlt{})} and (ii) \textbf{few-shot (\fsxlt{})} cross-lingual recommendation. Firstly, through \zsxlt{} we aim to investigate the capabilities of NNRs trained monolingually on English (i.e., on the MIND news) to generate recommendations in another language (i.e., in one of the 14 languages of \xmind{}). Under \zsxlt{}, the user history and candidates during training are \textit{monolingual}, in English only. Secondly, with \fsxlt{} we seek to determine whether target-language injection during training benefits the models' performance compared to pure \zsxlt{}. In the \fsxlt{} setting, we increasingly replace a portion (varying from 10\% to 90\%) of the English training set (both in history and candidate set) with target-language news. For a fair setup (i.e., no knowledge of test data distributions during training), the distribution of languages in our validation sets mirror the language ratios of respective training sets \cite{schmidt2022don}.

We couple the two training scenarios (monolingual and bilingual), with two corresponding types of \textit{news consumption patterns} during inference: (i) \textbf{monolingual} (denoted \texttt{MONO}) -- the user reads news and receives suggestions only in the target language, and (ii) \textbf{bilingual} (denoted \texttt{BILING}) -- the user consumes news in English and in another language, and recommendations are also provided in the same two languages. To construct the bilingual user history, and candidate set, respectively, we randomly replace a portion of the English news with corresponding \xmind{} translations in the target language. Like in bilingual training, we also vary the portion of replaced news in the interval $[10\%, 90\%]$, with a 10\% step. 

The two training setups, each combined with both consumption patterns, thus result in four types of experiments: \textbf{(i)} \textbf{\zsxltmono{}} -- monolingual training (in English) and evaluation on monolingual news consumption in the target language; \textbf{(ii)} \textbf{\zsxltbiling{}} -- monolingual training (in English) and evaluation on bilingual news consumption in English and the target language; \textbf{(iii)} \textbf{\fsxltmono{}} -- bilingual training in a mixture of English and target-language and evaluation on monolingual news consumption in the target language \textbf{(iv)} \textbf{\fsxltbiling{}} -- bilingual training in a mixture of English and target-language and evaluation on bilingual news consumption in English and the target language.\footnote{Note that due to the high computational costs, in the case of \fsxltbiling{} we replace the same percentage of English news with articles in the target language both during training and testing (e.g., 10\% RON with 90\% ENG during training results in the same mixture in testing).}

\newcolumntype{g}{>{\columncolor{Gray}}c}
\newcolumntype{a}{>{\columncolor{ashgrey}}c}

\begin{table*}
    \centering
    \caption{\zsxltmono{} recommendation performance. For each model, we report performance (i) on the English MIND dataset (denoted \texttt{ENG}), (ii) averaged across all 14 target languages in \xmind{} (denoted \texttt{AVG}), and (iii) the relative percentage difference between average \zsxltmono{} and \texttt{ENG} performance ($\%\Delta$). We report averages, standard deviations across three runs. The best results per column are highlighted in bold, the second best are underlined.}
    \label{tab:zs_xlt_mono}
    \resizebox{\textwidth}{!}{%
    \begin{tabular}{lcgacgacgacga}
        \toprule
        \multirow{2}{*}{\textbf{Model}}
        & \multicolumn{3}{c}{\textbf{AUC}} & \multicolumn{3}{c}{\textbf{MRR}} 
        & \multicolumn{3}{c}{\textbf{nDCG@5}} & \multicolumn{3}{c}{\textbf{nDCG@10}} 
        \\ \cmidrule(lr){2-4} \cmidrule(lr){5-7} \cmidrule(lr){8-10} \cmidrule(lr){11-13}

        & ENG & AVG & $\%\Delta$ 
        & ENG & AVG & $\%\Delta$ 
        & ENG & AVG & $\%\Delta$ 
        & ENG & AVG & $\%\Delta$ \\ \hline

        NAML\textsubscript{CAT}
        & \multicolumn{2}{c}{55.46$\pm$0.18}  
        & 0.0
        & \multicolumn{2}{c}{31.12$\pm$0.56}  
        & 0.0
        & \multicolumn{2}{c}{29.44$\pm$0.67}  
        & 0.0
        & \multicolumn{2}{c}{35.81$\pm$0.59}  
        & 0.0
        \\ \hdashline
        
        CAUM-PLM
        & \underline{57.82$\pm$3.01}  
        & 55.90$\pm$1.75 
        & -3.32 
        & 32.92$\pm$1.68  
        & 31.38$\pm$1.62  
        & \underline{-4.68} 
        & 31.09$\pm$1.88  
        & 29.60$\pm$1.76 
        & \underline{-4.77} 
        & 37.49$\pm$1.71  
        & 35.96$\pm$1.58  
        & -4.08 
        \\

        LSTUR-PLM
        & 56.80$\pm$1.36  
        & \underline{56.28$\pm$1.68} 
        & -0.92 
        & 33.00$\pm$0.59  
        & 31.53$\pm$0.85  
        & \textbf{-4.47} 
        & 31.18$\pm$0.54  
        & 29.70$\pm$0.92 
        & \textbf{-4.73} 
        & 37.45$\pm$0.54  
        & 36.03$\pm$0.85  
        & \textbf{-3.78} 
        \\

        MANNeR
        & 50.00$\pm$0.00  
        & 50.00$\pm$0.00 
        & \textbf{0.00} 
        & \underline{35.58$\pm$0.31}  
        & \underline{33.03$\pm$0.54}  
        & -7.15 
        & \underline{33.86$\pm$0.22}  
        & \underline{31.34$\pm$0.48} 
        & -7.45 
        & \underline{40.17$\pm$0.21}  
        & \underline{37.64$\pm$0.44}  
        & -6.28 
        \\

        MINER
        & 57.73$\pm$7.33  
        & 55.81$\pm$4.33 
        & -3.32 
        & 31.71$\pm$4.95  
        & 30.20$\pm$3.42  
        & -4.76 
        & 30.01$\pm$4.95  
        & 28.53$\pm$3.52 
        & -4.91 
        & 36.45$\pm$4.84  
        & 35.02$\pm$3.51  
        & \underline{-3.90} 
        \\

        MINS-PLM
        & \textbf{59.89$\pm$0.29}  
        & \textbf{56.94$\pm$1.40} 
        & -4.93 
        & 34.75$\pm$0.24  
        & 33.11$\pm$0.51  
        & -4.70 
        & 32.94$\pm$0.23  
        & 31.32$\pm$0.52 
        & -4.93 
        & 39.35$\pm$0.20  
        & 37.64$\pm$0.50  
        & -4.35 
        \\

        NAML-PLM
        & 52.85$\pm$2.27  
        & 52.49$\pm$2.60 
        & \underline{-0.68} 
        & \textbf{35.98$\pm$0.44}  
        & \textbf{33.98$\pm$0.95}  
        & -5.56 
        & \textbf{34.11$\pm$0.46 } 
        & \textbf{32.13$\pm$1.08} 
        & -5.80 
        & \textbf{40.43$\pm$0.39}  
        & \textbf{38.38$\pm$1.02}  
        & -5.06 
        \\

        TANR-PLM
        & 54.18$\pm$5.91  
        & 53.27$\pm$1.91 
        & -1.68 
        & 35.47$\pm$0.95  
        & 32.14$\pm$0.9  
        & -9.40 
        & 33.56$\pm$1.07  
        & 30.25$\pm$1.02 
        & -9.98 
        & 40.03$\pm$0.86  
        & 36.78$\pm$0.88  
        & -8.11 
        \\
        
    \bottomrule
    \end{tabular}
    }
\end{table*}

\section{Results and Discussion}
We first analyze the \zsxlt{} performance of the benchmarked models, comparing it with their performance on the English data (i.e., MIND). 
Then we examine the NNRs' capabilities in \fsxlt, to determine the extent to which having (some) training data in the target language influences the recommendation performance. 
An ideal NNR should rank positive candidates higher than negative ones regardless of their language. We thus focus our analysis mostly on the ranking performance (i.e., the nDCG@10 metric).

\subsection{Zero-Shot Cross-Lingual Transfer}

Table \ref{tab:zs_xlt_mono} shows the \zsxltmono{} recommendation performance of the models, averaged over the 14 languages of \xmind{}, in comparison to their recommendation results on MIND (i.e, when both trained and evaluated in English). To contextualize these findings, we additionally report the performance of NAML\textsubscript{CAT}, which embeds news based only on their topical categories, completely ignoring the content. As a content/language-agnostic recommender, NAML\textsubscript{CAT} is thus an apt baseline both in English and the target languages. 

Firstly, in English recommendation on MIND, all NNRs improve over the category-based recommender, with relative improvements in nDCG@10 ranging from 1.78\% (i.e., MINER) to 12.91\% (i.e., NAML-PLM). 
In \zsxltmono{}, however, NNRs exhibit weaker performance w.r.t. the content-agnostic NAML\textsubscript{CAT}, with at most 7.19\% relative gain (i.e., NAML-PLM), and in some cases even lower performance (e.g., MINER with 2.19\% relative drop). Secondly, we find that NNRs trained monolingually in English and evaluated on the target languages of \xmind{} suffer an average relative drop in performance between 3.78\% (i.e., LSTUR-PLM) and 8.11\% (i.e., TANR-PLM) w.r.t. to their English performance. Models that achieve the best (ranking) performance in English (e.g., NAML-PLM, MANNeR, MINS) still (i) perform best in \zsxltmono{}, exhibiting, however, (ii) the highest relative performance drops w.r.t. their English performance. 
While this might lead to the conclusion that such NNRs are less robust in XLT, one should keep in mind that a \textit{random} recommender would exhibit the same performance regardless of the language of the news (i.e., 0\% drop in XLT performance w.r.t. English). In other words, the absolute XLT performance of NNRs still matters more than the relative drops w.r.t. their English performance. 

\begin{figure}[t]
     \centering
     \includegraphics[width=\columnwidth]{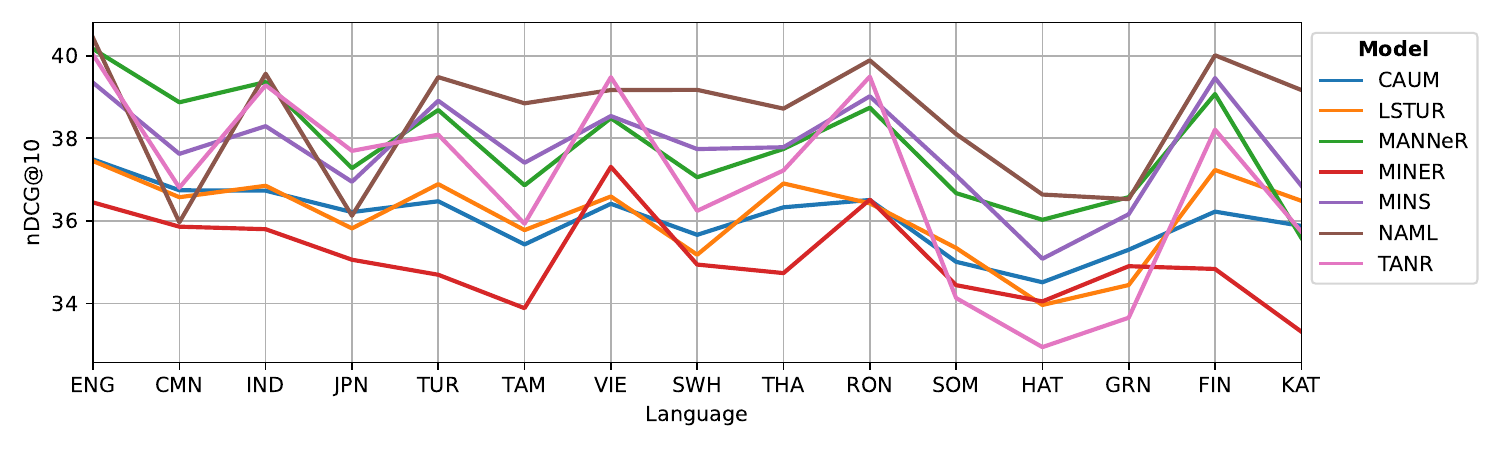}
     \caption{\zsxltmono{} ranking performance, w.r.t. nDCG@10, across the 14 languages in \xmind{} and English.}
     \label{fig:mono_train_mono_test_lang}
    \vspace{-0.5em}
\end{figure}

We next analyze the models' \zsxltmono{} performance across the 14 target languages individually, captured by Fig. \ref{fig:mono_train_mono_test_lang}. Expectedly, the NNRs achieve the best results on \texttt{VIE}, \texttt{IND}, \texttt{RON}, and \texttt{FIN}, which constitute some of the most represented languages in terms of numbers of tokens in the pretraining corpora of XLM-RoBERTa \cite{conneau2020unsupervised}. At the same time, we observe the lowest performance across models for \texttt{KAT}, \texttt{HAT}, and \texttt{GRN}, with the latter two languages being out-of-sample (i.e., not seen in pretraining) for XLM-RoBERTa. 

\begin{figure}[t]
     \centering
     \includegraphics[width=\columnwidth]{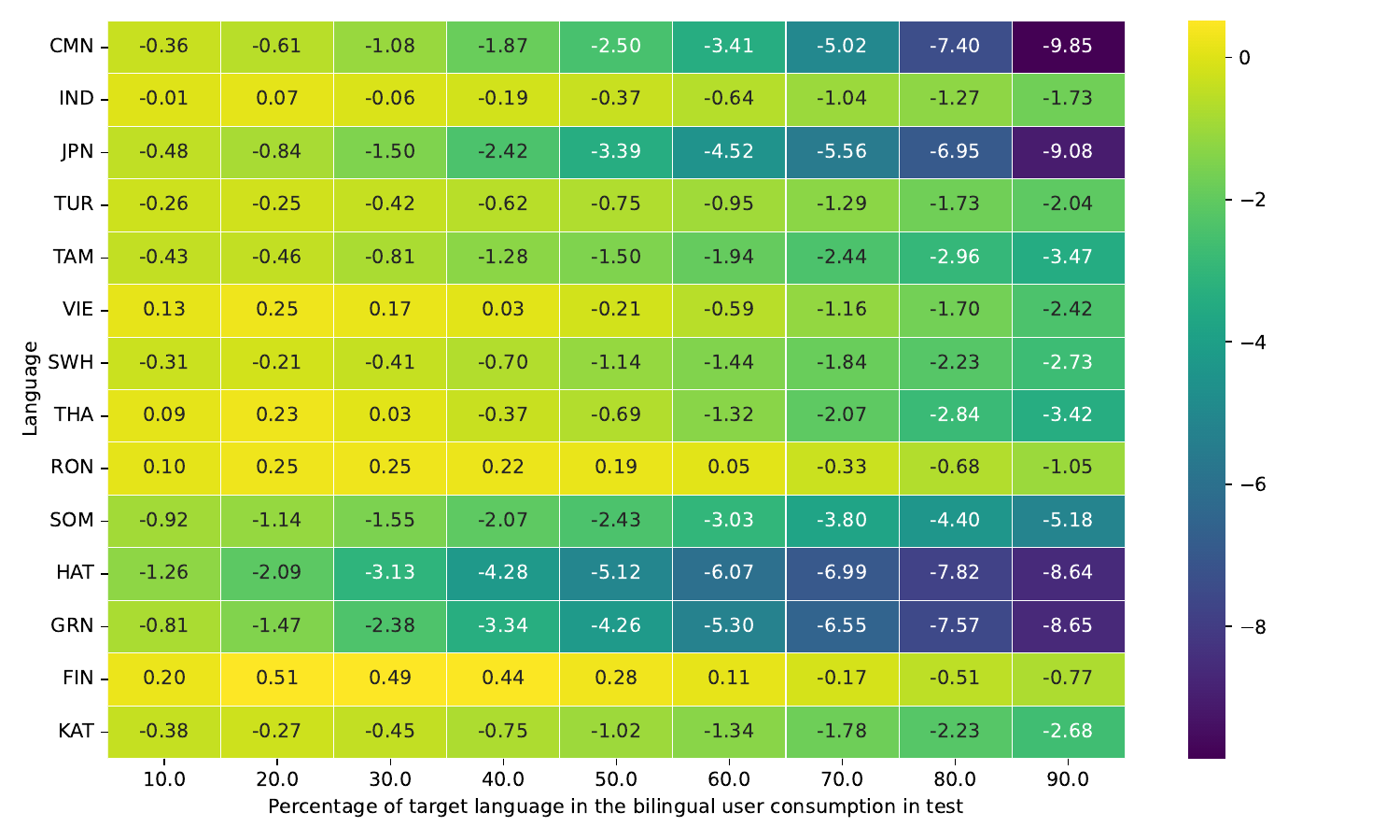}
     \caption{Relative percentage difference in ranking performance (w.r.t. nDCG@10), under \zsxltbiling{} compared to full English training and testing, for NAML-PLM.}
     \label{fig:mono_train_biling_test_naml}
    \vspace{-0.5em}
\end{figure}

\begin{figure*}[t]
     \centering
     \begin{subfigure}[b]{0.46\textwidth}
         \centering
         \includegraphics[width=\textwidth]{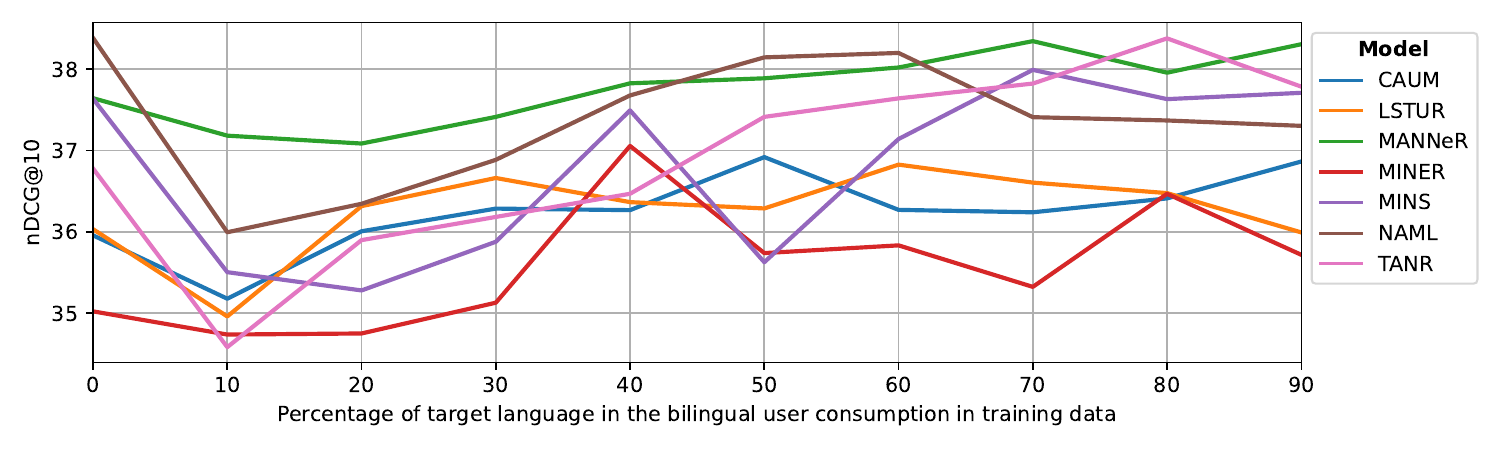}
         \caption{\fsxltmono{} setup.}
         \label{fig:biling_train_mono_test_avg}
     \end{subfigure}
     \hfill
     \begin{subfigure}[b]{0.46\textwidth}
         \centering
         \includegraphics[width=\textwidth]{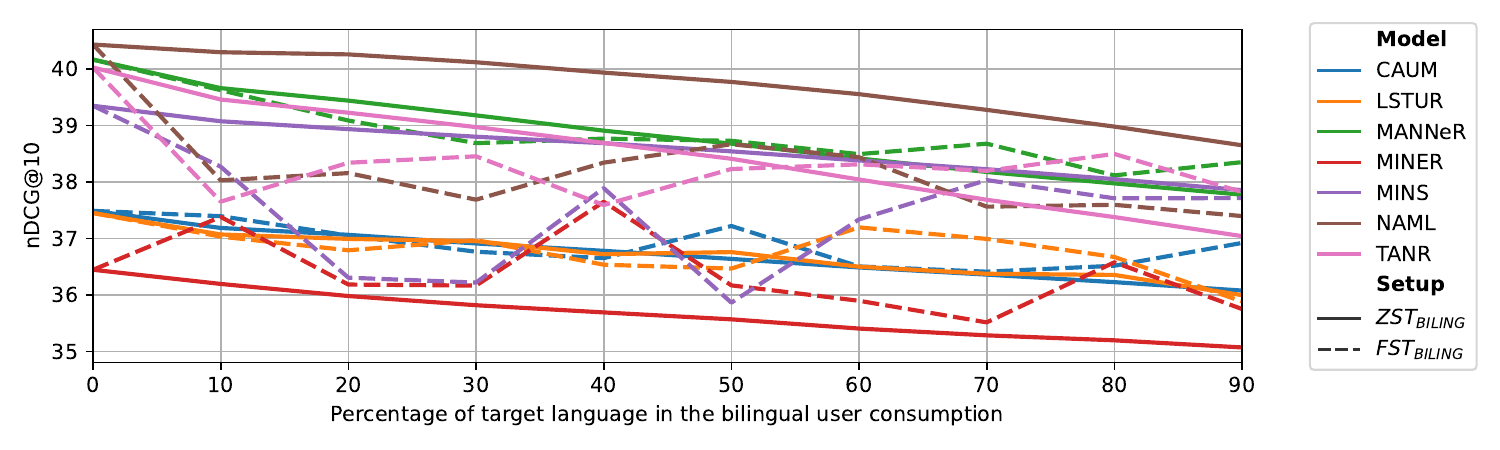}
         \caption{\fsxltbiling{} setup.}
         \label{fig:biling_train_biling_test_avg}
     \end{subfigure}
     \vspace{-0.5em}
    \caption{\fsxlt{} ranking performance, averaged over the 14 languages of \xmind{}, for various portions of target language in the user's bilingual news consumption.}
    \label{fig:fsxlt_results}
    \vspace{-0.5em}
\end{figure*}

\begin{figure*}[t]
     \centering
     \begin{subfigure}[b]{0.46\textwidth}
         \centering
         \includegraphics[width=\textwidth]{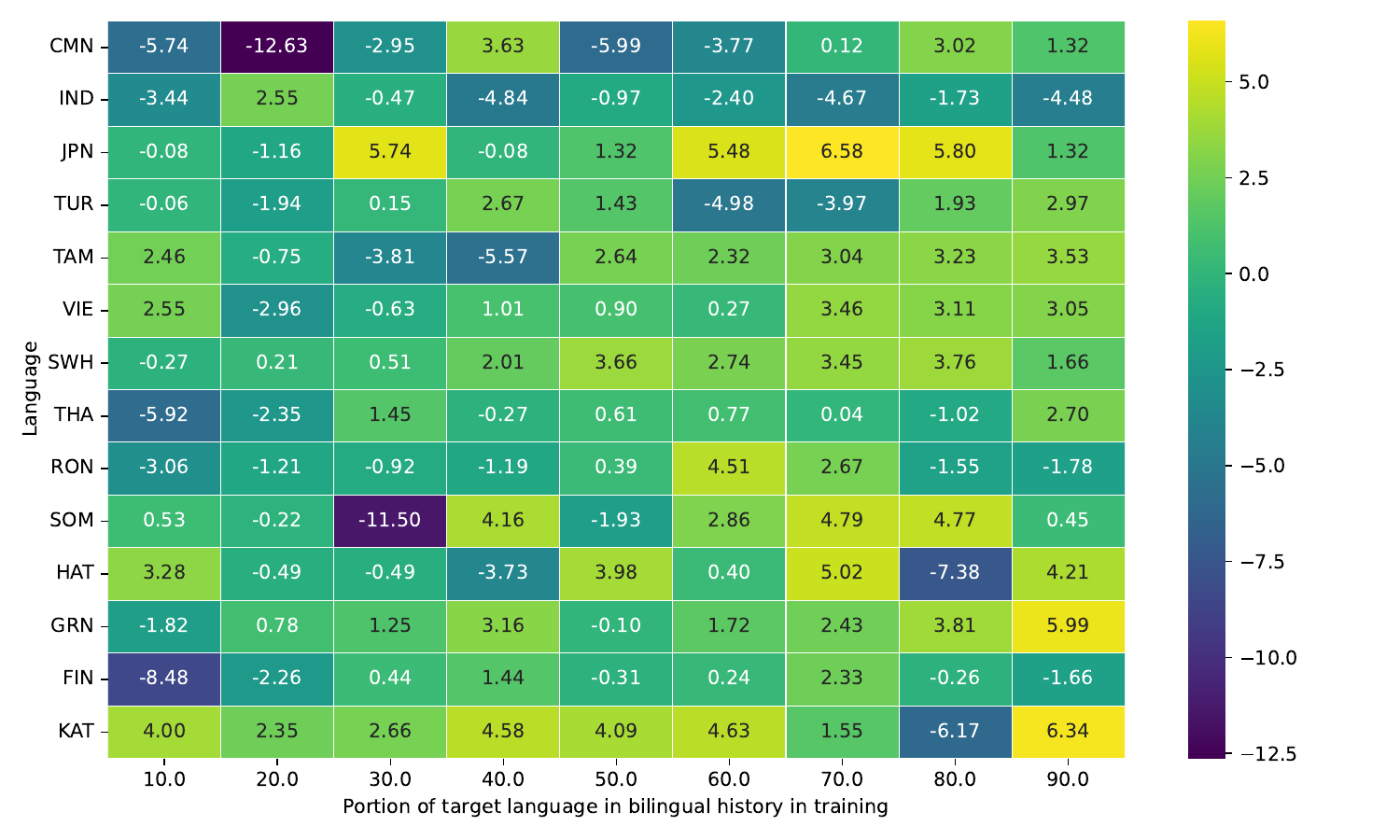}
         \caption{\fsxltmono{} compared to \zsxltmono{} ranking performance for MANNeR.}
         \label{fig:biling_train_mono_test_manner}
     \end{subfigure}
     \hfill
     \begin{subfigure}[b]{0.46\textwidth}
         \centering
         \includegraphics[width=\textwidth]{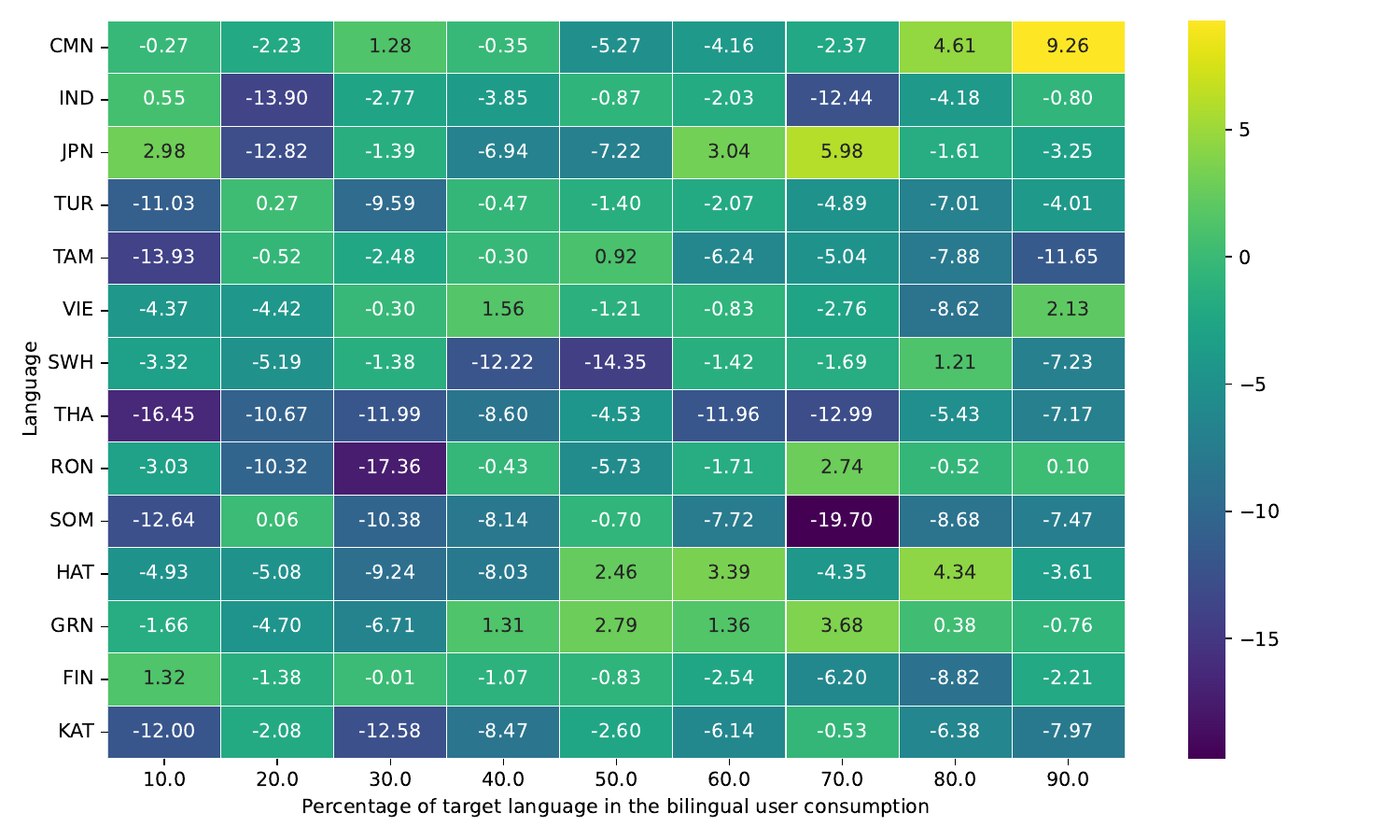}
         \caption{\fsxltbiling{} compared to \zsxltbiling{} ranking performance for NAML-PLM.}
         \label{fig:biling_train_biling_test_naml}
     \end{subfigure}
     \vspace{-0.5em}
    \caption{Relative percentage difference in ranking performance (w.r.t. nDCG@10), under \fsxlt{} compared to \zsxlt{}.}
    \label{fig:fsxlt_results_per_lang}
    \vspace{-0.5em}
\end{figure*}   

Next, we examine the change in the models' ranking performance when the user consumption is bilingual (i.e., \zsxltbiling{}). Fig. \ref{fig:mono_train_biling_test_naml} shows the results relative to the corresponding English performance for NAML-PLM (i) across target languages  and (ii) for varying percentages of the target language in the user's news consumption. Overall, we notice a steady decrease in performance for all models correlated with higher percentages of target language in the consumption pattern. Additionally, the performance of all NNRs deteriorates most drastically for the languages unseen by the mPLM during pretraining, namely \texttt{HAT} and \texttt{GRN}. Surprisingly, although based on the same mPLM, the NNRs are not equally robust to the choice of the target language. For example, NAML-PLM's performance drops up to 9.85\% when the user history contains \texttt{CMN} -- a higher decrease than for \texttt{HAT} or \texttt{GRN} -- whereas MANNeR's (relative) drop is at only 3.16\%. Additionally, we find that for some languages and models (e.g., LSTUR-PLM and \texttt{RON} or \texttt{FIN}), the decline in performance is lower when either English or the target language represent the predominant language in the user's news consumption. Given the models' diverse designs, this points to the need to investigate the robustness of the architecture, particularly of the UE, to changes in the composition of the user history, which so far, has only been assumed to be monolingual.

\subsection{Few-Shot Cross-Lingual Transfer}
Few-shot transfer, which requires the injection of a few target-language instances during model training, has been shown to yield sizable performance gains in NLP tasks \cite{schmidt2022don}. Therefore, it is often leveraged as an effective remedy to the dramatic performance drops suffered by multilingual models in \zsxlt{} setups, particularly for resource-lean target languages that are linguistically distant from the source language \cite{lauscher2020zero}. Motivated by these findings, we further analyze whether adding target-language data in training also benefits news recommenders: 
Fig. \ref{fig:biling_train_mono_test_avg} shows the NNRs' ranking performance (nDCG@10), averaged over all 14 target languages, for increasing percentages of target-language news included in the training data. Our results show that incorporating some target language data in training indeed ameliorates the performance losses from \zsxltmono{}. However, we find that if the target language constitutes less than 30-40\% of training news the recommendation performance drops (compared to \zsxlt{}). We believe that this is due to the relatively short user histories, consisting of 33 articles on average: a small percentage of news in another language is likely to confound the recommenders' UE. Moreover, although higher portions of target-language training data lead, on average, to higher gains over \zsxlt{}, the gains vary per model and percentage of target-language news injected. More specifically, models such as NAML-PLM or CAUM-PLM tend to produce less accurate rankings when the two languages seen during training are unevenly represented (e.g., for low and high portions of the target language). Delving deeper into per-language results (Fig. \ref{fig:biling_train_mono_test_manner}), we find that \fsxlt{} particularly benefits low-resource languages and languages unseen in pretraining of XLM-RoBERTa, as the mPLM on which the NE of all NNRs in our evaluation are based. 

In contrast to the \fsxltmono{} setting, few-shot target-language injection does not appear to be equally effective when we assume a bilingual news consumption of users during evaluation, that is, in the \fsxltbiling{} scenario. Fig. \ref{fig:biling_train_biling_test_avg} compares the ranking performance (nDCG@10) of NNRs in \fsxltbiling{} against their respective performance in \zsxltbiling{}, for varying proportions of target-language news included in the training. The majority of NNRs perform on par or better when a few instances in the target languages are seen during training. However, the performance of NAML-PLM -- generally the best performing NNR in our evaluation -- is subpar to that achieved by the model trained only on English. A closer look at its performance across languages, shown in Fig \ref{fig:biling_train_biling_test_naml}, reveals that \fsxlt{} benefits the model primarily for those languages for which it exhibits the highest losses under \zsxltbiling{}, namely \texttt{CMN}, \texttt{JPN}, \texttt{HAT}, and \texttt{GRN}  (i.e., Fig. \ref{fig:mono_train_biling_test_naml}). For other NNRs, the gains are more evenly distributed across all languages. 

Overall, these variations in the performance of models and the limited benefits of few-shot target-language injection -- particularly when considering bilingual news consumption -- emphasize again the need for a deeper understanding of the specific factors that drive multilingual NNR performance in order to inform design of user encoder architectures that are robust to multilingual user histories. 
\begin{figure}[t]
     \centering
     \includegraphics[width=\columnwidth]{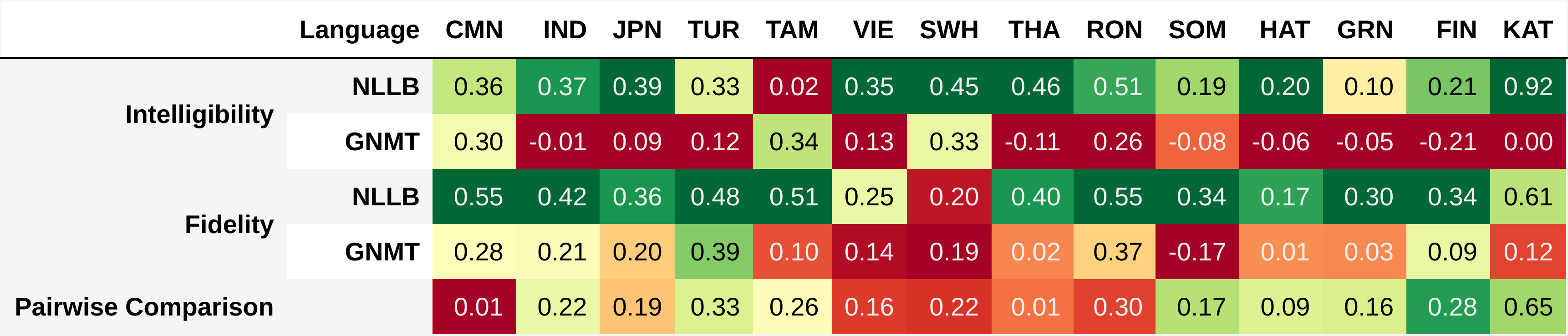}
     \caption{Annotator agreement in terms of Krippendorf's alpha, per language, for all questions in the annotation task.}
     \label{fig:annotator_agreement}
    \vspace{-0.5em}
\end{figure}

\section{Translation Quality}

We finally investigate the quality of the translations in \xmind{}. Concretely, we (i) estimate the translation quality and (ii) investigate the robustness of the NNRs to different translations of the source news from MIND \cite{wu2020mind}. To this end, we use \xmind{}small and additionally translate its training and development portions using Google (Neural Machine) Translation (GNMT) \cite{wu2016google}, a commercial MT system that supports all \xmind{} languages. 
GNMT has been shown to outperform NLLB on translation from English to various low-resource languages \cite{costa2022no}.
\footnote{GNMT is a proprietary system, hindering fair comparisons to open-source models due to the lack of transparency regarding its model architecture and training procedures.}
\footnote{We translated the text with the Cloud Translation - Advanced (v3) API: \href{https://cloud.google.com/translate/docs/overview}{https://cloud.google.com/translate/docs/overview}, using Google Cloud research credits worth approximately \$5,000.}

\subsection{Manual Quality Estimation of Translation}

Given the size of the \xmind{}small dataset and our (limited) annotation budget, it was infeasible to manually post-edit the translations of the news in the test portion of \xmind{}. We therefore resorted to conducting an annotation task to estimate the quality of the translations. To this end, we sample 50 news from the development portion of the MIND dataset (i.e. the portion used as test set in all our experiments), according to the (i) the distribution of categories in the dataset and (ii) the distribution of the total length of the news (i.e., composed of title and abstract). This way, we ensure that the sampled instances are representative of the full dataset. 

We carry out the annotations using the Potato annotation tool \cite{pei2022potato}. 
Two annotators judged the quality of the NLLB and GNMT translations for each language.\footnote{All annotators were native speakers of the target language and fluent in English. Most of the annotators were certified interpreters/translators of the target language.} 
The task comprised of a total of five questions, targeting three aspects of the translations: \textit{intelligibility}, \textit{fidelity}, and \textit{pairwise comparison} between the NLLB and GNMT translations. The first two questions were repeatedly asked independently for the NLLB and GNMT translations. The annotators answered the following questions: \textit{(1-2) Is the translation acceptable?} -- binary answer; \textit{(3-4) To which extent is the information from the original text accurately retained in the translation?} -- 5-point Likert-scale answers, ranging from "Not at all" to "Completely"; \textit{(5) Which translation is better?} -- categorical answer with three options, namely "Translation A", "Translation B", or "They are comparably good". In order to remove any position bias in all questions, we randomized the source of translations A and B shown to the annotator, such that 50\% of the time translation A stemmed from NLLB, and the remaining 50\% from GNMT. Overall, across most of the target languages, we observed higher annotators agreement (Krippendorf's alpha \cite{krippendorff2013content}) for the NLLB translations for the first two questions, as shown in Fig. \ref{fig:annotator_agreement}, than for GNMT, where we observe little to no agreement between the annotators.

\begin{figure}[t]
     \centering
     \begin{subfigure}[b]{\columnwidth}
         \centering
         \includegraphics[width=\columnwidth]{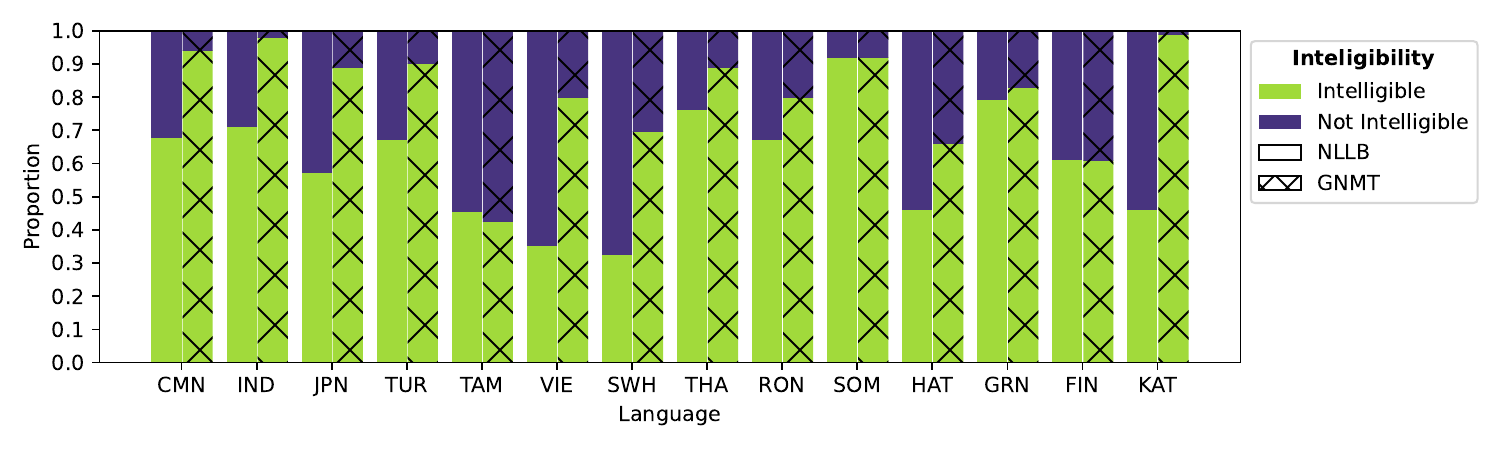}
         \caption{Intelligibility.}
         \label{fig:intelligibility}
     \end{subfigure}
     \vfill
     \begin{subfigure}[b]{\columnwidth}
         \centering
         \includegraphics[width=\columnwidth]{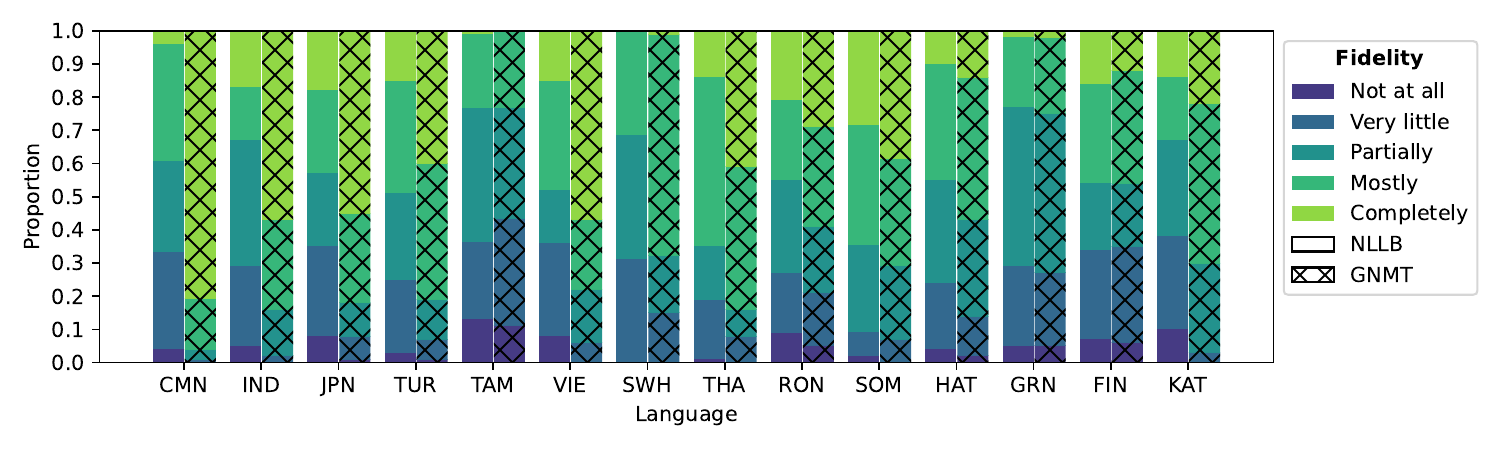}
         \caption{Fidelity.}
         \label{fig:fidelity}
     \end{subfigure}
     \vfill
    \begin{subfigure}[b]{\columnwidth}
         \centering
         \includegraphics[width=\columnwidth]{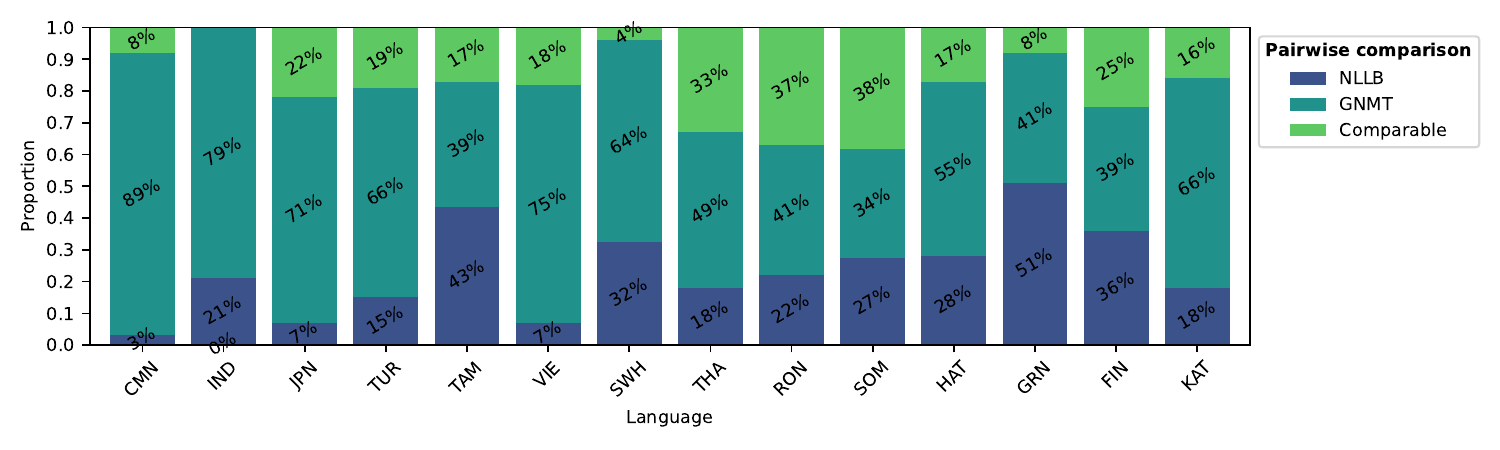}
         \caption{Pairwise comparison.}
         \label{fig:pairwise_comparison}
      \end{subfigure}
    \vspace{-0.5em}
    \caption{Annotation task results for each question type.}
    \label{fig:intelligibility_fidelity}
    \vspace{-1em}
\end{figure}

For over half of the target languages in \xmind{}, the annotators deemed the NLLB translations to be intelligible in at least 60\% of the cases (see Fig. \ref{fig:intelligibility}). Similarly, we find that our translations retain the information of the original texts, at least partially, in the majority of cases, as illustrated in Fig. \ref{fig:fidelity}. Notably, the NLLB-sourced translation are deemed more faithful to the original news than the GNMT-based ones particularly for low-resource languages such as \texttt{TAM} and \texttt{GRN}. This finding is corroborated by the results from the pairwise comparison, shown in Fig. \ref{fig:pairwise_comparison}, which show that NLLB translations are judged to be overall better than their GNMT counterparts for these two languages. Nonetheless, across all languages and aspects of evaluation, translations obtained with the commercial GNMT are deemed generally of higher quality than those generated with the open-source NLLB.  

\begin{figure}[t]
     
\end{figure}

The annotators' feedback revealed several challenges that contributed to the generally low scores assigned to both kinds of translations. Firstly, we remark that often one part of the news (e.g., title) was perfectly translated, whereas the other portion (e.g., abstract) was not accurately depicted in the target language. Secondly, in few cases, the phrasing of the news title was hard to comprehend even in English, impeding translation. Lastly, we note that given the US origins of the news articles from the MIND dataset \cite{wu2020mind}, many of the topics discussed in the news are not usually encountered in some of the target languages or they pertain solely to the US (e.g., sports news). In such cases, we observed that the MT systems performed particularly poorly. A closer look at the translation errors reveals that, in terms of intelligibility, both NLLB and GNMT generate worse translations for news in categories such as entertainment (e.g., for languages such as \texttt{VIE}, \texttt{TUR}, \texttt{JPN}), movies, music, or television (e.g., particularly for lower-resource languages \texttt{TAM} and \texttt{KAT}). Such errors can be explained by the fact that these categories of news tend to contain terms that exhibit more idiomaticity (e.g., especially in movie titles), which is well-documented source of trouble for MT \cite{dankers2022can}.  
Similar patterns emerge when analyzing the news categories on which the fidelity of the translations is lower. However, our results indicate that there is not a particular category on which one of the MT systems is better than the other (according to our annotators). Lastly, we observe that translations of shorter texts, obtained with both NLLB and GNMT, are of higher quality than those of longer ones. This can be explained by the fact that, at least NLLB, has been trained on pairs of shorter documents \cite{costa2022no}.

\subsection{Robustness of NNRs to Translation Quality}
We next investigate the robustness of the best-performing NNRs from our previous experiments, namely NAML-PLM and MANNeR, to translations obtained with different MT systems. To this end, we re-run the previous experiments for the version of the dataset translated with GNMT and compare the results against the ones obtained using the NLLB-translated \xmind{}small.

\begin{figure}[t]
     \centering
     \includegraphics[width=\columnwidth]{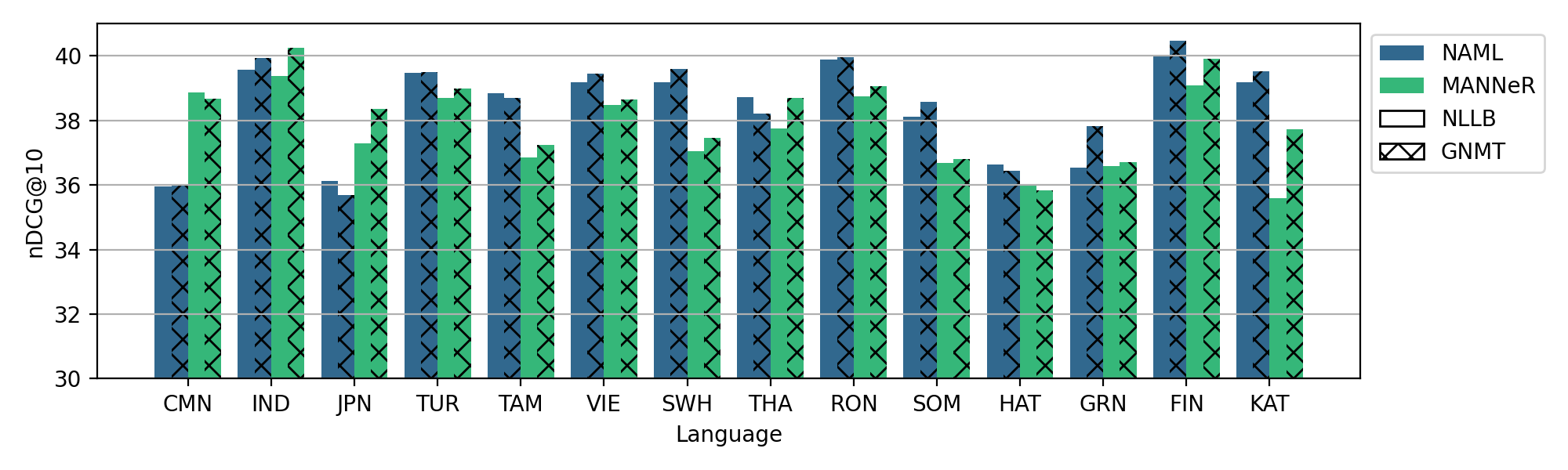}
     \caption{\zsxltmono{} ranking performance, w.r.t. nDCG@10, in terms of MT system, for NAML-PLM and MANNeR.}
     \label{fig:ablation_mono_train_mono_test_lang}
    \vspace{-0.5em}
\end{figure}

Fig.\,\ref{fig:ablation_mono_train_mono_test_lang} shows the ranking performance (w.r.t. nDCG@10) for both models and MT systems under \zsxltmono{} (i.e., training on the English MINDsmall and evaluation on the target language). The performance of both NNRs seems largely unaffected by the provenance of the translations. The differences are insignificant according to an independent samples T-test (for a p-value of 0.05), with the exception of MANNeR's performance on \texttt{IND} and \texttt{KAT}, where GNMT test translations lead to better ranked recommendations. This is important as it indicates that, although the GNMT translations were judged to be better on average than the ones produced by NLLB, the NNRs appear to be robust to differences in translation quality.

\begin{figure}[t]
     \begin{subfigure}[b]{\columnwidth}
         \centering
         \includegraphics[width=\columnwidth]{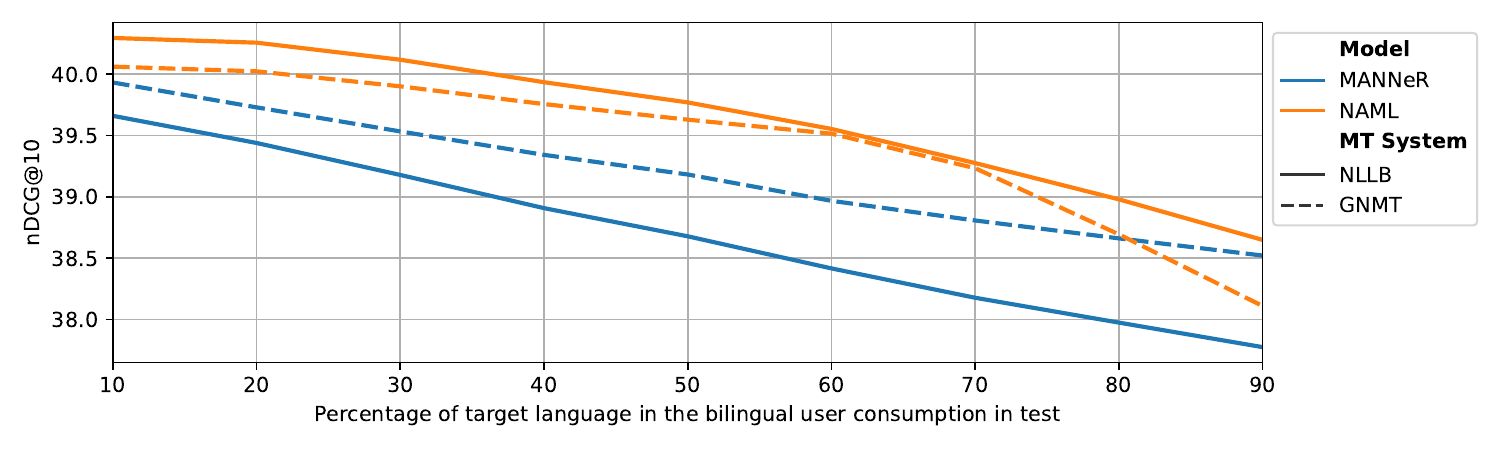}
         \caption{\zsxltbiling{} ranking performance.}
         \label{fig:ablation_mono_train_biling_test_avg}
     \end{subfigure}
     \vfill
     \begin{subfigure}[b]{\columnwidth}
         \centering
         \includegraphics[width=\columnwidth]{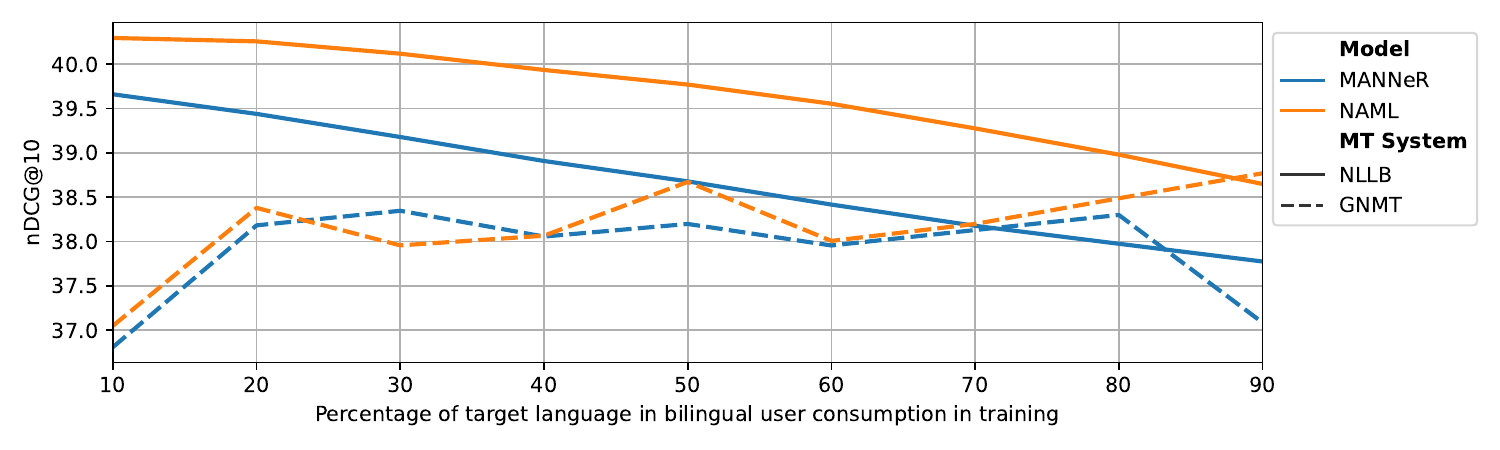}
         \caption{\fsxltmono{} ranking performance.}
         \label{fig:ablation_biling_train_mono_test_avg}
     \end{subfigure}
    \caption{Ranking performance, w.r.t. nDCG@10, in terms of MT system, for NAML-PLM and MANNeR.}
    \label{fig:ablation_avg}
    \vspace{-2em}
\end{figure}

We further check this hypothesis in the \zsxltbiling{} setting. The corresponding results from Fig. \ref{fig:ablation_mono_train_biling_test_avg} demonstrate small differences in performance depending on the MT system used. However, the differences are again not statistically significant according to the same independent T-test. We observe similar patterns and no statistical significance in both \fsxlt{} settings (e.g., Fig \ref{fig:ablation_biling_train_mono_test_avg} illustrates the ranking performance for the \fsxltmono{} case). Overall, these findings indicate that, despite the infeasibility of manual post-editing of the test set translations in \xmind{}, the quality of the translations obtained with the open-source NLLB (i) is on par with those generated by a state-of-the-art commercial MT system, and (ii) has no significant effect on the NNR's recommendation performance.

\section{Conclusion}

The ever-growing language-diversity of online news readers has not been reflected in the news recommendation research, which focuses nearly entirely on resource-rich languages, particularly English, and monolingual news consumption. In this work, we introduced \xmind{}, an open multilingual news recommendation dataset comprising 14 linguistically and geographically diverse languages, derived from the English MIND dataset using machine translation. We used \xmind{} to benchmark several state-of-the-art content-based NNRs in both zero-shot and few-shot cross-lingual recommendation transfer, experimenting with both monolingual and bilingual news consumption patterns. Our findings show that current NNRs suffer considerable performance losses under \zsxlt{}, while the inclusion of target-language data in \fsxlt{} training brings limited gains to recommenders, especially in the context of bilingual news consumption. We believe that \xmind{} is a valuable resource for the news recommendation community, and hope it will foster much more research on multilingual and cross-lingual news recommendation, for speakers of both high- and low-resource languages.
\begin{acks}
The authors acknowledge support by the state of Baden-Württemberg through bwHPC and the German Research Foundation (DFG) through grant INST 35/1597-1 FUGG. This material is based upon work supported by the Google Cloud Research Credits program with the award EDU275608761.
\end{acks}

\bibliographystyle{ACM-Reference-Format}
\balance
\bibliography{references}

\end{document}